\documentclass[slac_one]{revtex4}
\usepackage{graphicx}
\usepackage{fancyhdr}
\begin{document}

\newcommand{\drbar}{{\overline{DR}}}
\newcommand{\msbar}{{\overline{MS}}}

\title{{\small{2005 International Linear Collider Workshop - Stanford,
U.S.A.}}\\ 
\vspace{12pt}
SUSY Studies} 

%

\author{Jan Kalinowski}
\affiliation{Institute of Theoretical Physics, Warsaw University, Warsaw,
  Poland }

\begin{abstract}
This report summarizes the progress in SUSY
 studies performed since the 
last International Linear Collider Workshop in Paris (LCWS04). 

\end{abstract}

\maketitle

\thispagestyle{fancy}


\section{INTRODUCTION} 
The Standard Model provides a very good and economical description for
all experimental data. However, there is a number of key physics 
questions that the Standard Model (SM) is not able to address, e.g.: 
\begin{itemize} 
\item 
What is the origin of mass? Is the Higgs mechanism,  or its 
  variants, behind the gauge symmetry breaking?\\[-7mm]
\item
What is the origin of matter-antimatter asymmetry?\\[-7mm]
\item 
What are the properties of neutrinos?
  \footnote{Neutrinos provide   
the first  experimental evidence for physics beyond the SM.}\\[-7mm]
\item 
Do all forces, including gravity, unify?\\[-7mm]
\item 
What is the nature of dark matter, dark energy?
\end{itemize}
What is interesting and stimulating for our Working Group is  that
supersymmetry (SUSY) may turn out to be related to all these
questions. Moreover, SUSY can experimentally be tested at future colliders:
Large Hadron Collider (LHC) and International Linear Collider (ILC). 
In particular, the  ILC may provide
essential tools for discovery answers. Discovering supersymmetry would mean a
grand revolution in particle physics.

\begin{figure}
(a) \hspace{4cm} (b) \hspace{4cm} (c) \hspace{3.5cm} (d)\\[-2mm]
\includegraphics[width=4.2cm, height=4.1cm]{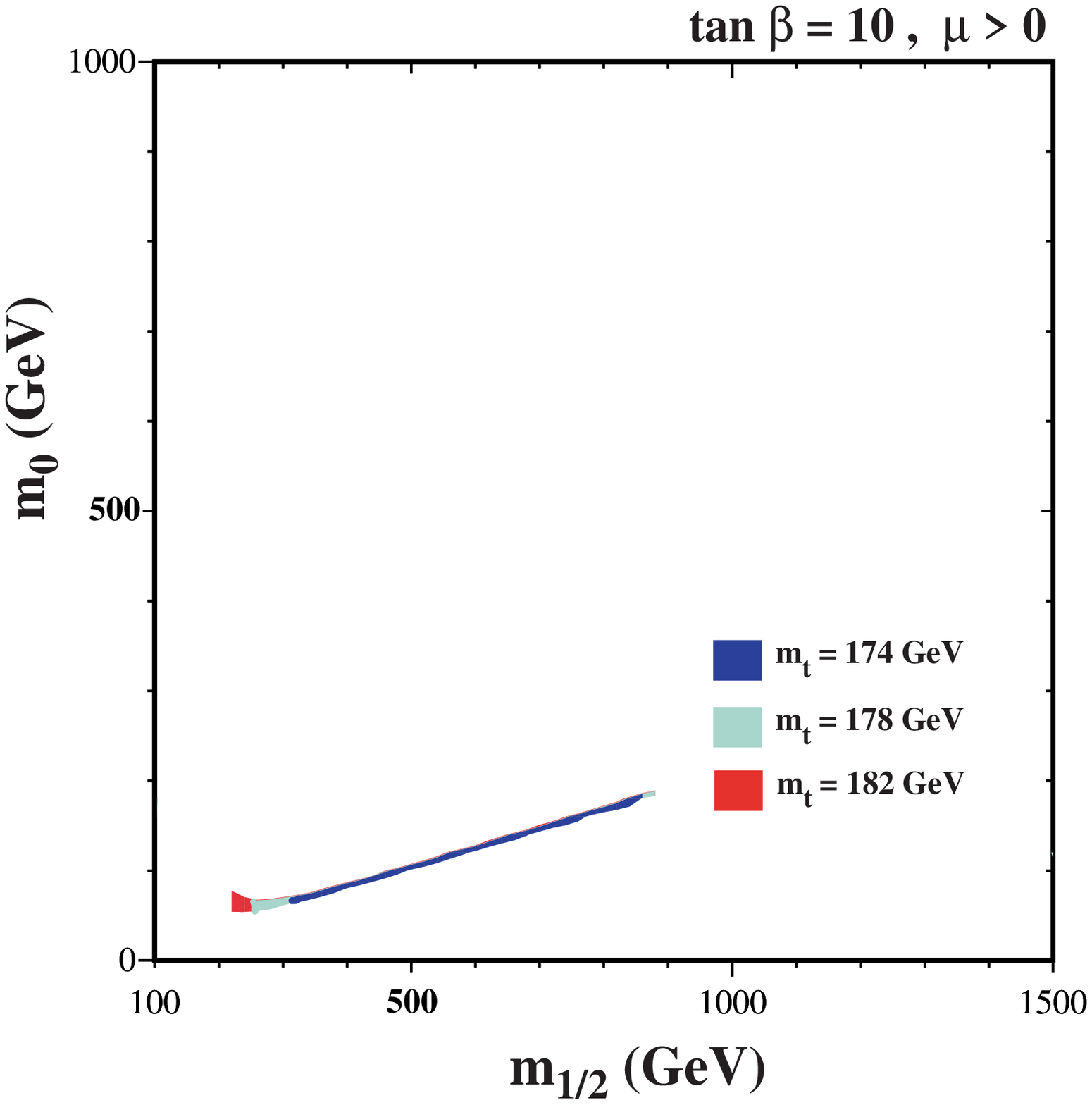}
\includegraphics[width=4.6cm,height=4.2cm]{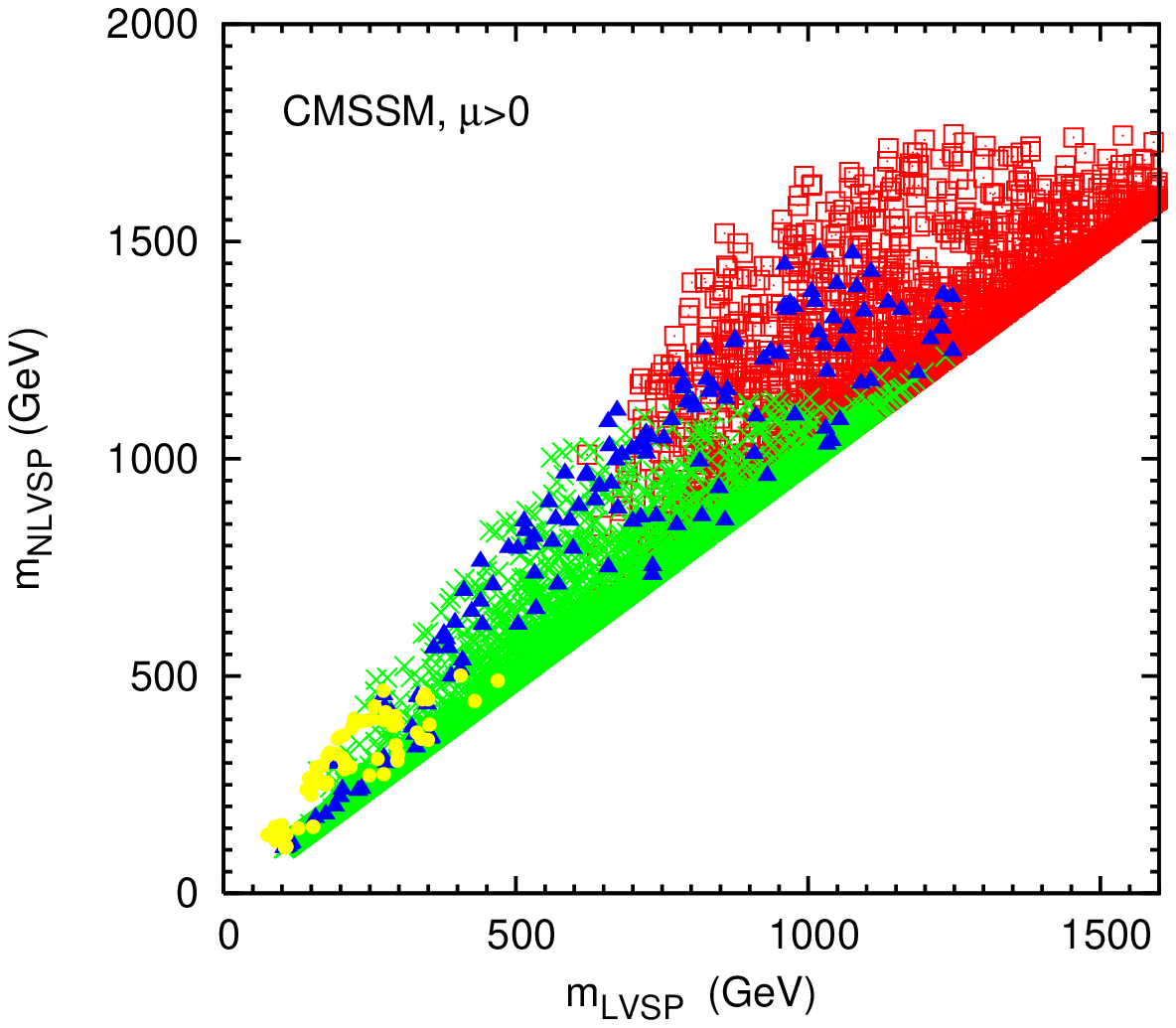}
\includegraphics[width=4.7cm, height=4cm]{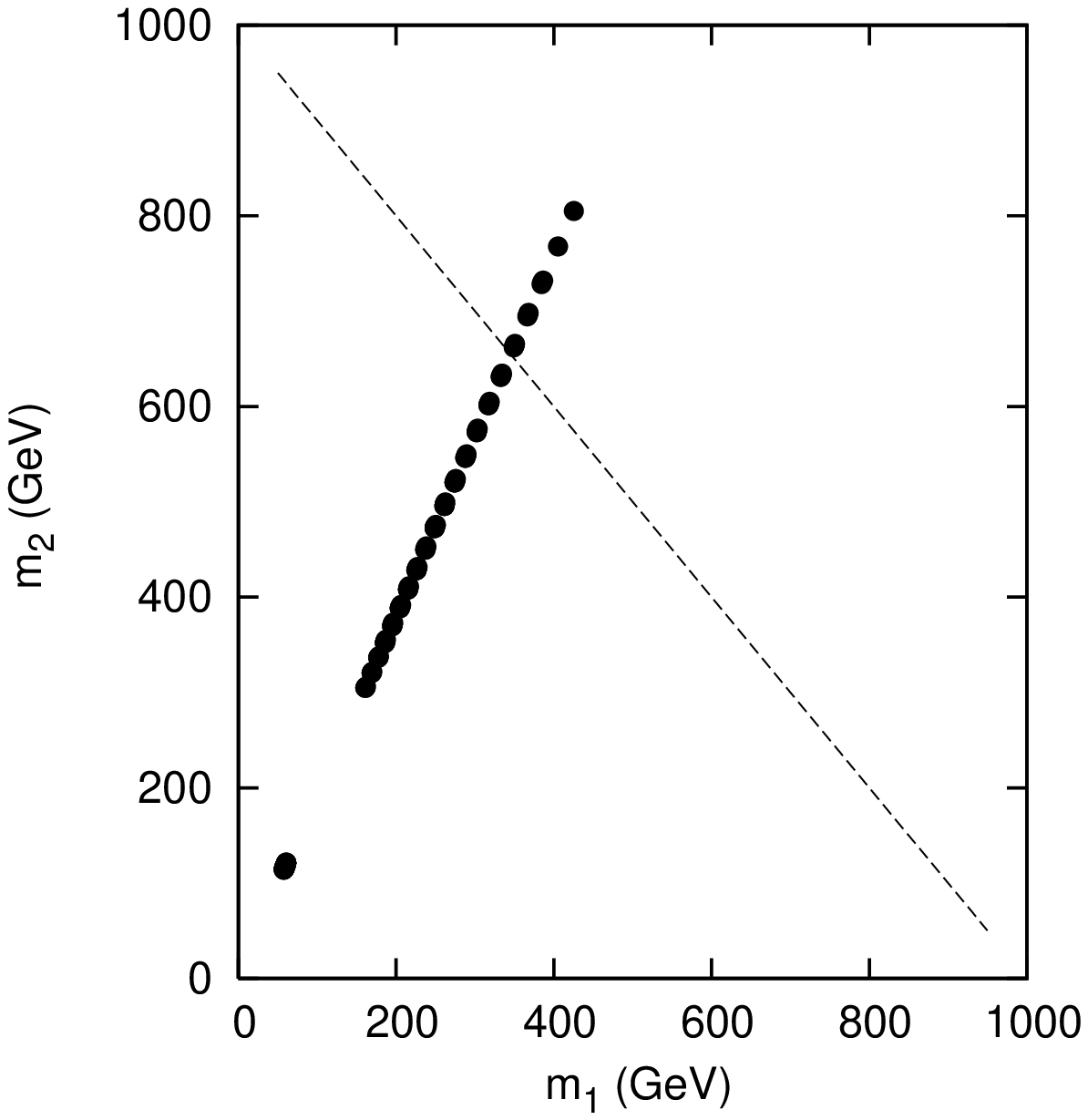}
\includegraphics[width=3.9cm, height=3.8cm]{fit.eps}
\caption{(a) The WMAP strip for $\mu > 0$, $A_0 = 0$
and $\tan\beta = 10$ in the cMSSM; from  \cite{Heinemeyer:2004xw}. 
(b) Scatter plot of the masses of the LVSP
and the NLVSP in the cMSSM; from \cite{Ellis:2004bx}. 
(c) Scatter plot of the masses
$m_1$=$m_{\tilde{\chi}^0_1}$ and $m_2$=$m_{\tilde{\chi}^0_2}$ of 
cMSSM models satisfying
$\tan\beta$ = 10, 20, 35 and $A_0=0$. 
(d) $\chi^2$ fits based on the current
experimental EW  precision observables and DM constraints 
as functions of $m_{1/2}$ in the cMSSM for different $A_0$, $\tan\beta = 10$; 
from \cite{Heinemeyer:2004xw}.}
\label{fig:cmssm}
\end{figure}

Why supersymmetry is so attractive? Technically, of the many motivations for
the supersymmetric   extension of the SM, 
perhaps the most important, next to the connection to
gravity,  is the ability to  stabilize the electroweak scale and predict the
gauge unification. With R-parity conserved the lightest superparticle (LSP), 
in many models the lightest neutralino $\tilde{\chi}^0_1$, is a
candidate for the main constituent of cosmological cold dark matter (DM).   
In fact the data from WMAP and astrophysics/cosmology \cite{WMAP}
can already put constraints  on
many possible supersymmetric models. For example, within the
constrained MSSM (cMSSM - often referred to as mSUGRA) with universal scalar
($m_0$) and  gaugino ($m_{1/2}$) masses and universal trilinear ($A_0$) 
scalar couplings at some
unification scale, the dark matter 
data (and low-energy and collider constraints) 
select a thin strip in the $m_0$-$m_{1/2}$
parameter plane, shown in fig.\ref{fig:cmssm}(a),  in which $\mu>0$, 
$A_0=0$ and $\tan\beta=10$ \cite{Heinemeyer:2004xw}. A scan of the
cMSSM  
parameter space gives a broader band, as shown in the plane of the 
lightest visible (LVSP) and the  next-to-lightest visible (NLVSP) 
superparticle masses, see fig.\ref{fig:cmssm}(b); the LSP 
itself was considered not to be visible \cite{Ellis:2004bx}.  
This figure gives an impression that most of the points (in this specific
model) are outside the reach of the ILC.  However, at the ILC in most cases  
the lightest neutralino $\tilde{\chi}^0_1$ 
will clearly be detected and its mass measured. Therefore 
a better representation of the reach of the
ILC is shown in fig.\ref{fig:cmssm}(c)   in which a large fraction of
the allowed combinations of parameters
scanned for the previous plot would give signals
accessible to ILC, at least through
$e^+e^- \to \tilde{\chi}^0_1\tilde{\chi}^0_2$ \cite{KP}. 
The  isolated spot at low
neutralino masses  is from the focus
point region of parameter space (also omitted in \cite{Ellis:2004bx}). 

From this discussion it follows that 
SUSY can naturally be compatible with present  electroweak (EW) 
measurements. In fact, 
the fit to EW and DM data, again within the cMSSM,
points to rather low values of SUSY breaking parameters, as seen in
fig.\ref{fig:cmssm}(d) \cite{Heinemeyer:2004xw}, 
which interestingly enough are  close to the benchmark point
SPS1a of \cite{Allanach:2002nj}.     

However, nothing tells us that the cMSSM is the right
framework (most probably it is not). 
As discussed during the meeting  
the parameter space
widens very much when model assumptions are relaxed and many 'crazy'
scenarios are possible \cite{Baer}.  For example, if the Higgs and
scalar mass parameters at the GUT scale   are different, 
$m^2_{H_u}=m^2_{H_d}\neq m^2_0$, 
two solutions for  relic density are found: 
one is neutralino annihilation via heavy Higgs resonance
even at low values of $\tan\beta$,  
while the other is neutralino annihilation via higgsino components
at low values of $m_0$   \cite{Baer:2005bu}. Breaking the $m^2_{H_u}=m^2_{H_d}$
relation opens the light squark/slepton co-annihilation regions etc. 
This indicates how important will be to determine
SUSY breaking parameters from future 
experimental measurements with minimum number of
theoretical assumptions. Proving SUSY will require not only to discover new
particles and  
measure  their masses, decay widths and production cross sections,
but also \\[1mm]
\phantom{m}~~$\bullet$ verify that they are superpartners, {\it i.e.}~measure
their spin and parity, gauge quantum numbers and couplings,\\ 
\phantom{m}~~$\bullet$
 reconstruct the low-energy SUSY breaking parameters 
without assuming a specific scenario,\\
\phantom{m}~~$\bullet$ and ultimately unravel the SUSY breaking mechanism 
sheding light on physics at high (GUT?, Planck?) scale.\\[1mm]
Since the last 
International Linear Collider Workshop in Paris (LCWS04)
members of the SUSY working groups were very 
active delivering many talks during three  
regional meetings: the ALCPG
Workshop in 
Victoria, the ECFA Workshop in Durham,  and the ACFA Workshop in Taipei; 
the transparencies can be found
in \cite{meetings}. Here in Stanford (LCWS05)
we had 21 presentations in the SUSY WG session
alone, and a number of SUSY related talks in other sessions. It is impossible
to give justice to all in my talk (and these proceedings) and I apologize for
omissions.

\section{ACTIVITIES}

The aim of our activities is to strengthen the physics case for the  $e^+e^-$
ILC  demonstrating that it   
would be an indispensable tool. Our activities  also include studies of
synergy of the LHC and the ILC showing that    
coherent analyses of data from the LHC {\it and} ILC would allow for a 
better, model independent reconstruction of low-energy  SUSY
parameters, and  connect  low-scale phenomenology with the high-scale
physics.  
The main themes of our last year's activities, among other things,  include:
setting the SPA convention and project, 
refined experimental simulations and analyses,    
higher-order theoretical calculations, 
LHC/ILC synergy, 
non-minimal scenarios with  CP, lepton flavour,
and/or R$_p$ violation, cosmology connection etc.


\subsection{SPA Convention and Project}
Since the experimental accuracies expected at ILC are at the 
per-cent down to the per-mille level \cite{Weiglein:2004hn,R3B}, 
the future experimental effort  must be met by equally precise theoretical
computations. The implementation of higher-order corrections calls for 
a well defined framework for the calculational schemes in
perturbation theory as well as for the input parameters. The proposed 
{\it Supersymmetry Parameter Analysis: Convention and Project} \cite{SPA} 
provides 
\begin{itemize}
\item 
SPA Convention for SUSY  Lagrangian parameters and  SM input parameters,
\\[-7mm]
\item 
program repository with computer codes, \\[-7mm] 
\item  
tasks of the SPA Project with long- and short-term sub-projects,\\[-7mm]  
\item 
reference Point SPS1a' as a testbed for testing the SPA Project. 
\end{itemize}

SPA Convention adopts the $\drbar$  scheme 
\cite{Siegel:1979wq,Jack:1994rk} for the SUSY Lagrangian
parameters.  It is based on regularization by 
dimensional reduction together with modified
minimal subtraction. It has been shown in the meeting 
 \cite{Stockinger:2005gx} that the 
inconsistencies of the original scheme \cite{Siegel:1980qs} 
can be overcome and that
the $\drbar$ scheme can be formulated in a mathematically consistent way.
To make use of the highly developed
infrastructure for proton colliders, which is based on the $\msbar$ 
factorisation scheme 
(which requires {\it{ad-hoc}} counter terms to
restore supersymmetry), a finite shift from the commonly used $\msbar$ 
parton
density functions to the $\drbar$ density functions has to be carried out
\cite{37a}.   Moreover, for massive final state particles
spurious density functions for the 4-D gluon
components have to be introduced to comply with the
factorization theorem. 
Formulating an efficient combination of the most attractive elements of
both schemes in describing hadronic processes is therefore an important task 
of the project \cite{inprog}. 
The SUSY scale is chosen $\tilde M = 1$~TeV to avoid large 
threshold corrections in running the mass parameters by renormalization group
techniques from the high scale down to the low scale. 
In the decay widths and production cross sections 
the physical on-shell masses are introduced such that the phase-space is
treated in the observables closest to experimental on-shell kinematics; 
the masses of the light particles 
can generally be neglected in high energy processes. 

The program repository contains codes for
translations between different computational schemes, spectrum calculators,
codes for decay widths, cross sections, low-energy and
cosmological/astrophysical observables, event generators, RGE codes
etc. It is 
an open system and the responsibility for all programs 
remains with the authors. It is understood that in each case the theoretical
state-of-the-art 
precision is implemented. For communication between codes SLHA
\cite{Skands:2003cj}  
is strongly recommended, which is extended in a suitable way where appropriate.
SPA provides the translation tables and the links to the computer codes on
the SPA web page.

The tasks of the SPA Project aim at higher-order calculations, better
understanding of $\drbar$/$\msbar$ connections, improvements of experimental
and theoretical precision, investigations of LHC/ILC synergy, cold dark
matter, developments and explorations of beyond the MSSM scenarios etc.  Since
the goal of the SPA Project is to reconstruct the fundamental structure of the
supersymmetric theory at the high scale, the precise understanding and the
combination of all information, that will become available from collider and  
low-energy experiments and astrophysical/cosmological observations, will be
required.

The reference point SPS1a', a slight modification of SPS1a point to be 
consistent with all available experimental data,
is proposed as a testing ground to explore the potential of such extended
experimental and theoretical effort. Preliminary studies have shown that while
the ultimate aim of the SPA project can be achieved, additional work both on 
the theoretical as well as on the experimental side is still needed. In
particular, 
SPA should include detailed analyses of other benchmark points and
SUSY scenarios.

The SPA Project is a dynamical system expected to
evolve continuously. The current status of the Project,
listing the conveners responsible for specific tasks as
well as the links to the available calculational tools, can
be found at the SPA home page: www://spa.desy.de/spa.

\subsection{Need for higher-order calculations}

The present level of theoretical calculations 
still does not match the expected experimental precision,
particularly in coherent LHC+ILC analyses. For example, table~\ref{tab:scale} 
shows a crude estimate of lower limits on the theoretical errors in deriving
the superparticle masses in the SPS1a' point by shifting the
SUSY scale $\tilde{M}$ from 1 TeV down to 100 GeV.
While the experimental precision at LHC can be matched in general, 
another order-of-magnitude improvement 
is required in the theoretical precision to match the
expected experimental errors at ILC. 

\begin{table}[t] 
\begin{center}$
\begin{array}{|c|c|c||c|c|c|}
\hline
\mbox{Particle} & \mbox{Mass [GeV]} & \delta {\rm  ~[GeV]} 
& \mbox{Particle} & \mbox{Mass [GeV]} & \delta {\rm ~[GeV]} \\
\hline 
h^0     & 115.4 & 1.3 & H^0     & 431.1 & 0.7 \\ 
\tilde{\chi}^0_1   &  97.75& 0.4 & 
\tilde{\chi}^0_2   & 184.4 & 1.2 \\ 
\tilde{\chi}^\pm_1   & 183.1 & 1.3 & \tilde{\tau}_1 & 107.4 & 0.5 \\ 
 \tilde{e}_R    & 125.2 & 1.2 &
 \tilde{e}_L    & 190.1 & 0.4 \\
 \tilde q_R     & 547.7 & 9.4  &
 \tilde q_L     & 565.7 & 10.2 \\
 \tilde t_1   & 368.9 & 5.4  &
 \tilde b_1   & 506.3 & 8.0  \\  
\hline
\end{array}$
\end{center}
\caption{ Superparticle masses for the SPS1a' Reference Point with the 
         SUSY scale  $\tilde{M} =$ 1 TeV, 
         and their variation when $\tilde{M}$ is shifted down to 100 GeV; 
         from \cite{SPA}.}  
\label{tab:scale}
\end{table}

\begin{figure}
\includegraphics[width=4cm, height=4cm]{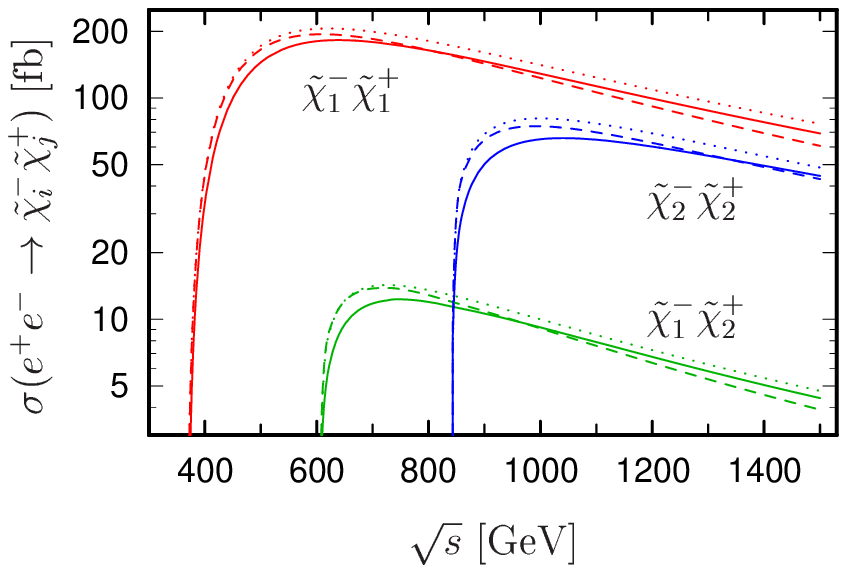}
\includegraphics[width=4cm, height=4cm]{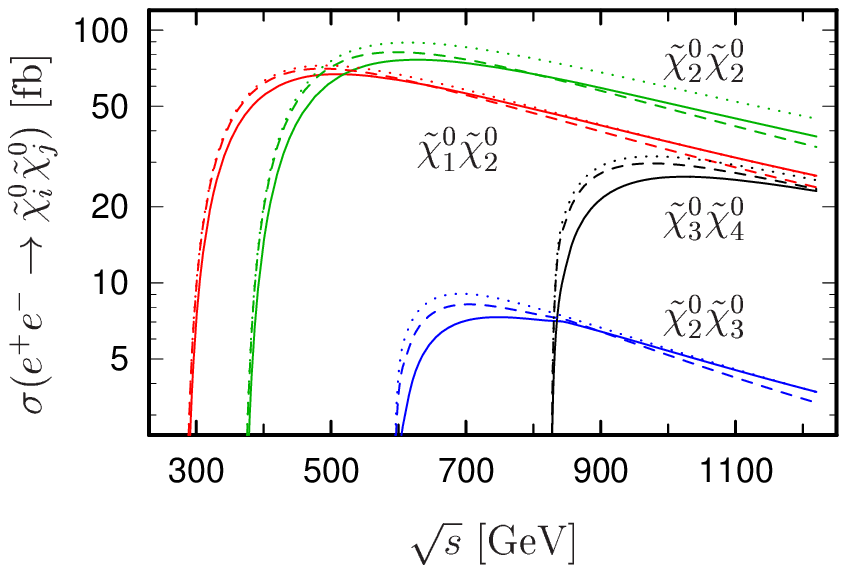}
\includegraphics[width=4cm, height=4cm]{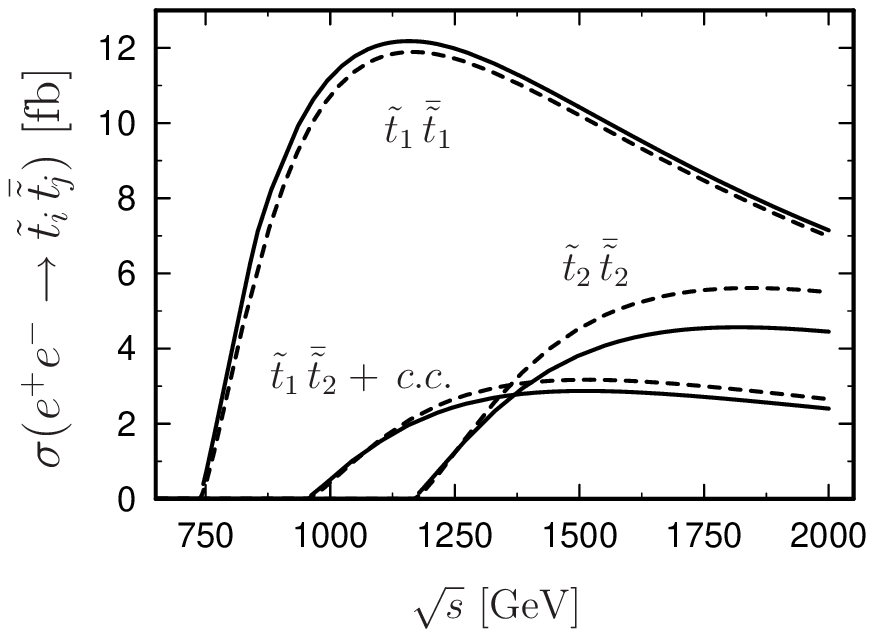}
\includegraphics[width=4cm, height=4cm]{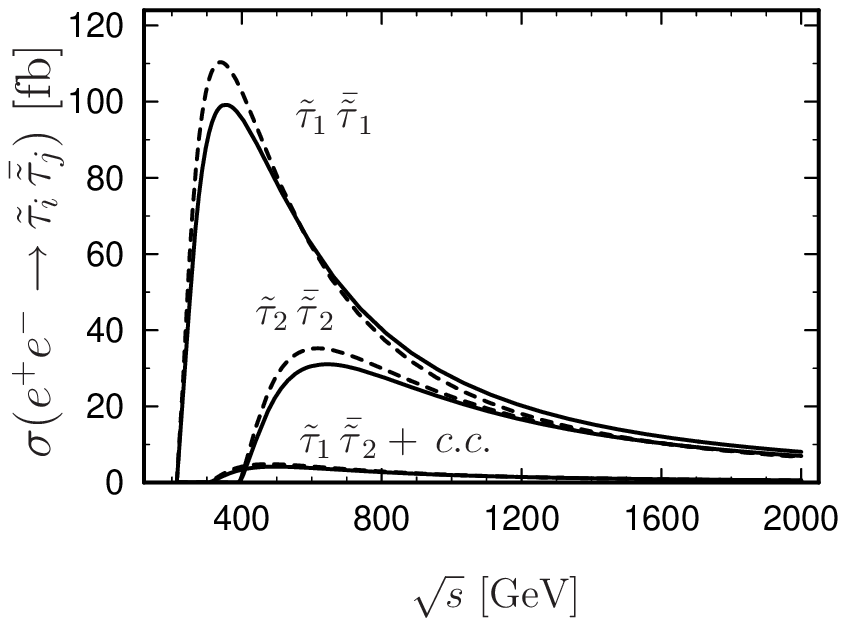}
\caption{Chargino, neutralino, stop and stau production cross sections for the
  Reference Point SPS1a'; from \cite{Kovarik:2005sk}.}
\label{fig:HO}
\end{figure}

Recent high precision calculations for production processes of the
SUSY particles at the ILC have been reviewed by K.~Kova\v{r}\'{i}k
\cite{Kovarik:2005sk}.  The input parameters of SPS1a' benchmark point 
have been translated to on-shell which then have been used as input in the
calculation of the pair production of
stops, sbottoms, staus, charginos and neutralinos in $e^+ e^-$
collisions. The advantage of using the on-shell input values is that the
well-established procedure of on-shell renormalization can be
applied. Plots in fig.~\ref{fig:HO} show 
total cross-sections of the complete ${\cal O}(\alpha)$, 
leading higher-order and tree-level results for the chargino, neutralino, 
stop and stau production processes. 
For neutralino and chargino production the separation of the
QED and weak corrections followed the prescription of the SPA 
project. 

\subsection{Experimental analyses}

Masses of superparticles can be measured very precisely in threshold scans
\cite{R3B}, which however need running time. It is therefore important
to measure the masses in the continuum above thresholds including 
beam/bremsstrahlung effects \cite{beambrem}.  To achieve a
precision measurement one has to suppress efficiently SM background
processes  as well as disentangle signals coming simultaneously from different
SUSY channels.  Equally important is to verify the chirality assignment of
sfermions. 

\begin{figure}[t]
\begin{minipage}[b]{10cm}
(a)\includegraphics[width=4cm, height=3cm]{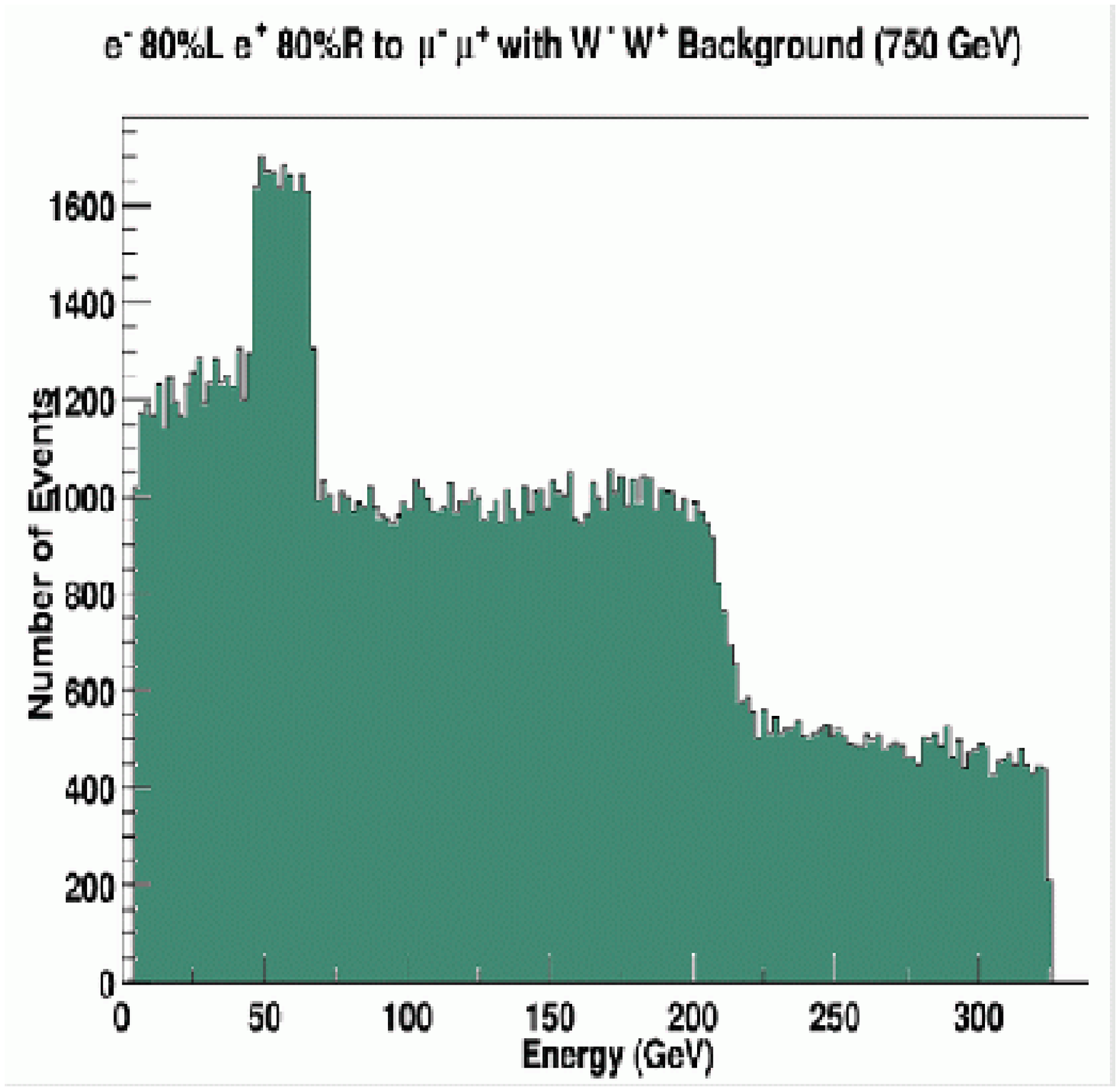}
(b)\includegraphics[width=4cm, height=3cm]{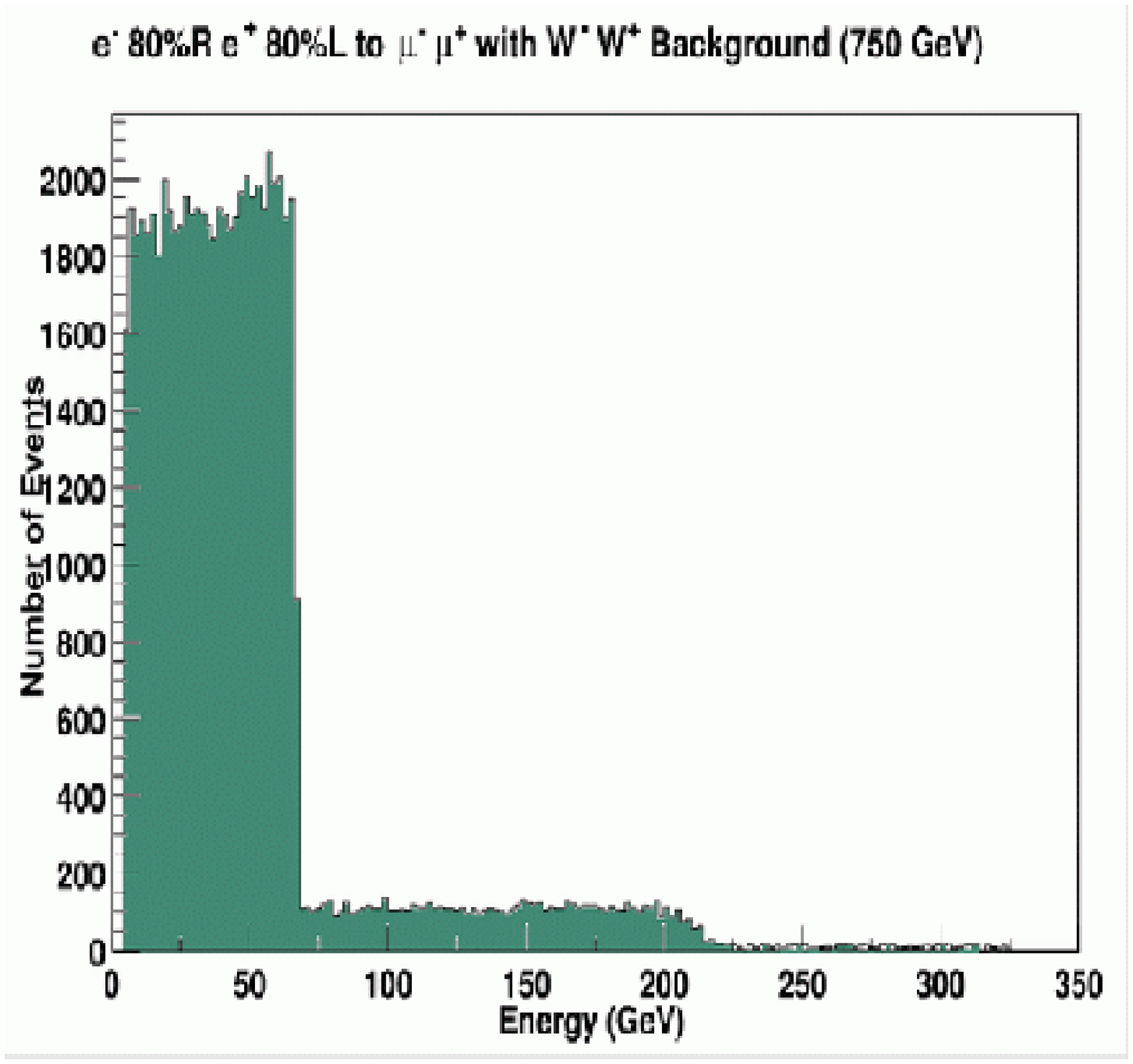}~~(c)
\par\vspace{2.2cm}
\end{minipage}\hspace{-15mm}
\begin{minipage}[t]{4cm} 
\includegraphics[width=7cm, height=6cm]{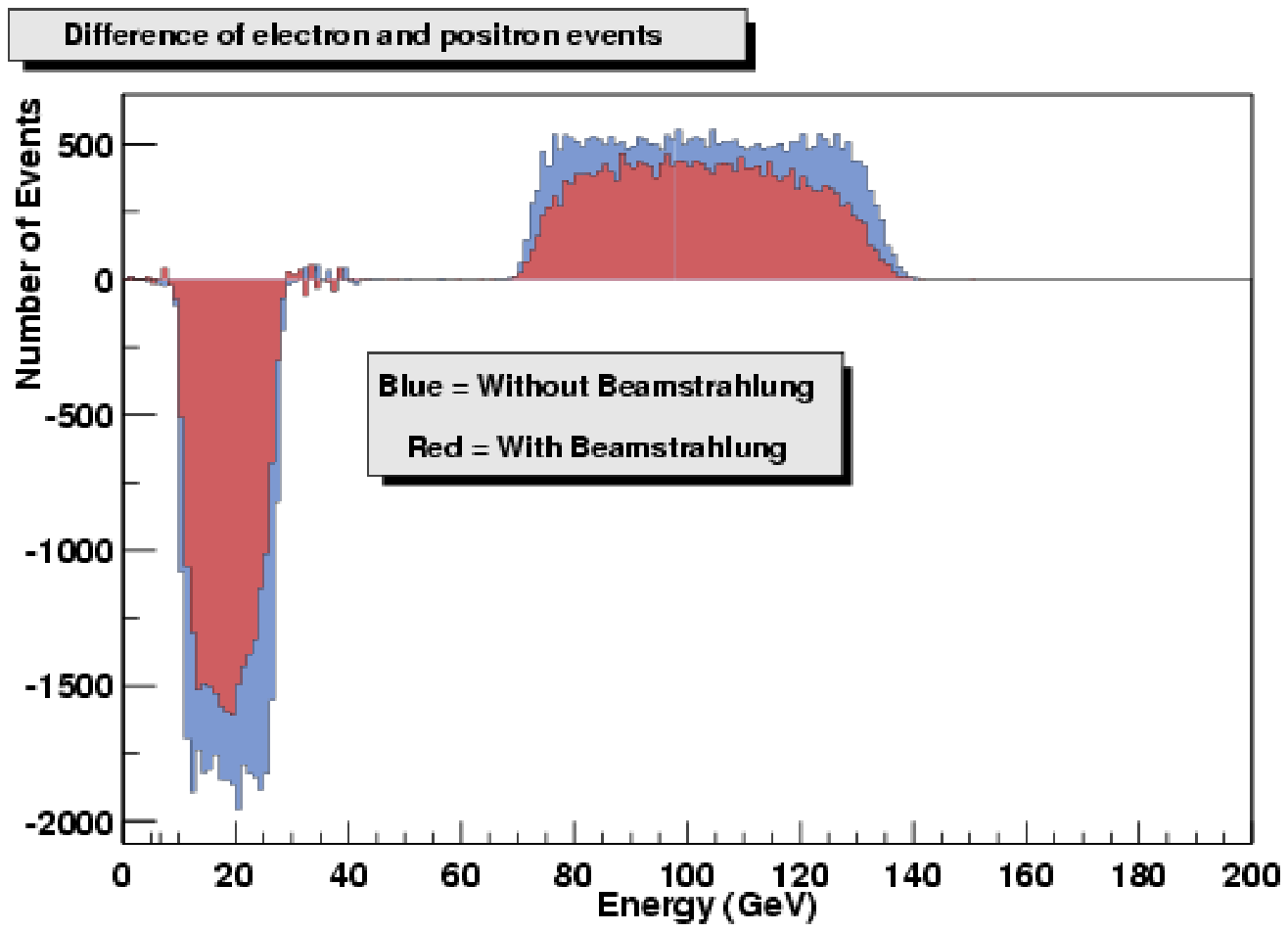}
\end{minipage}
\vspace{-2cm}
 \caption{(a,b) Energy spectra 
of muons from 
$\tilde{\mu}_{L,R}$ 
decays into $\mu\tilde{\chi}^0_1$ including 
$W^+W^-$ background for two combinations of beam polarisations  
$P_{e^-},P_{e^+}$: (a) $-$80\%,+80\%,  (b) +80\%,$-$80\%). 
(c) Difference of $e^+$ and $e^-$ energy  spectra from selectrons 
$\tilde{e}_{L,R}$ produced with polarised beams; from \cite{nauenberg}}.  
\label{fig:pol}
 \end{figure}

In this respect the ability of having both beams, positrons
and electrons, polarised is particularly important
\cite{Moortgat-Pick:2005cw}. In some scenarios  even 
 100\% electron polarisation may prove
insufficient to disentangle 
$\tilde{e}_{\rm L}^+\tilde{e}_{\rm R}^-$ and $\tilde{e}_{\rm R}^+
\tilde{e}_{\rm R}^-$ pairs and to test chiral quantum numbers. 
For smuon mass
measurement  the worst background coming from $W^+W^-$ final states
can easily be suppressed with right-handed electron/left-handed positron
beams, as shown in fig~\ref{fig:pol}(a,b) from \cite{nauenberg}. The 
selectron mass measurement, thanks to larger production 
cross sections, can greatly be 
improved  by a double subtraction of
$e^+$ and $e^-$ energy spectra and opposite electron beam polarizations, see
fig.~\ref{fig:pol}(c).   In both cases the endpoints from $\tilde{\mu}_{L,R}$  
and $\tilde{e}_{L,R}$ are clearly exposed giving rise to precise determination
of slepton masses. 
New techniques have been also developed \cite{Gerbode:2005ke} to isolate 
from SM backgrounds the selectron-decay signal in the forward region
($|\cos\theta|>0.8$) which in  some scenarios carries most of the information
constraining the selectron mass. With new selection techniques it has been 
 demonstrated that the
selectron signal can be separated from Standard Model 
backgrounds through the entire forward tracking region
$|\cos \theta| < 0.994$.

\begin{figure}
\begin{minipage}[b]{13cm}
(a)\includegraphics[width=3.5cm, height=3cm]{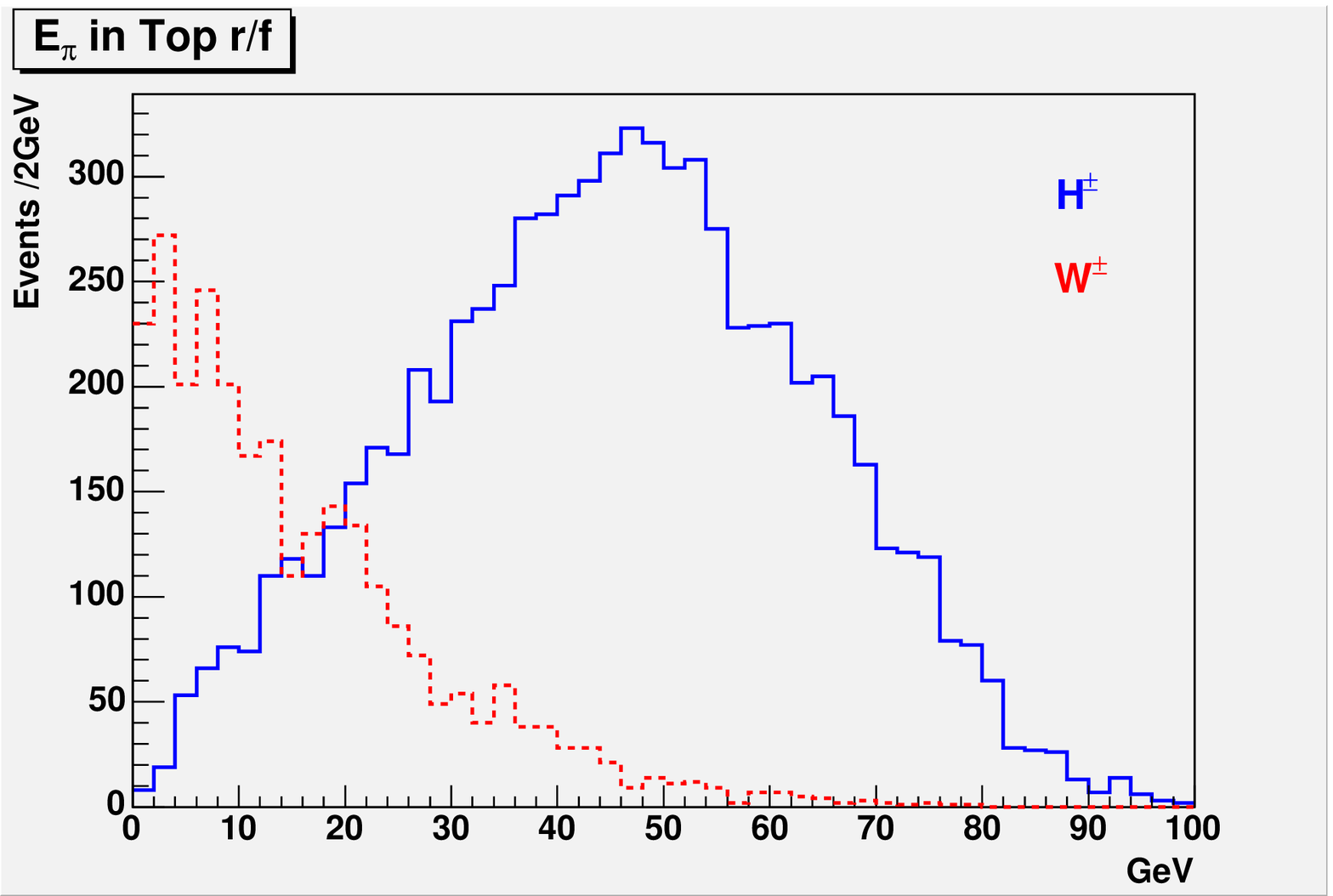}
(b)\includegraphics[width=3.5cm, height=3cm]{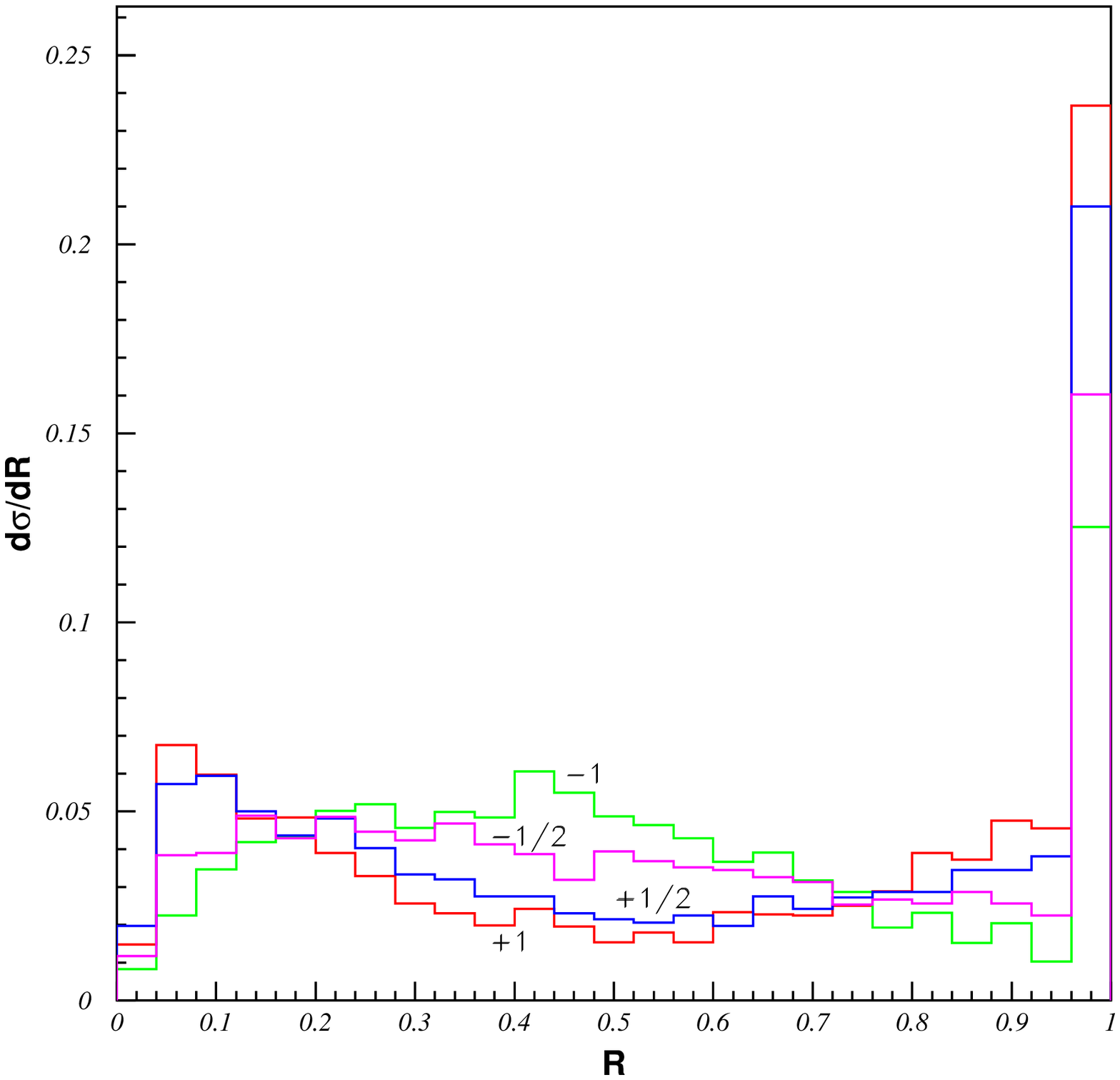}
(c)\includegraphics[width=3.5cm, height=2.5cm]{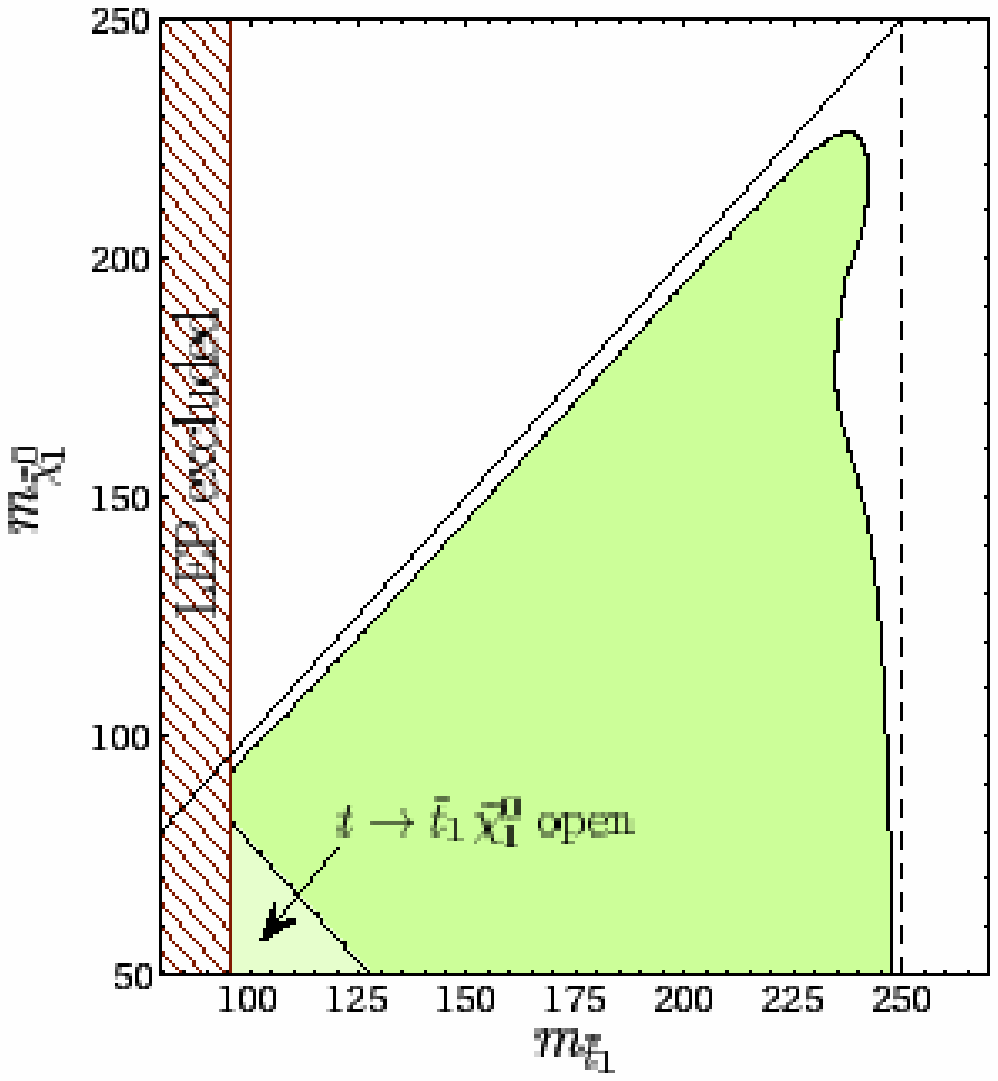}~~(d)
\par\vspace{-33mm}
\end{minipage}
\begin{minipage}[b]{3cm} 
\includegraphics[width=3.3cm, height=3.5cm,angle=-90]{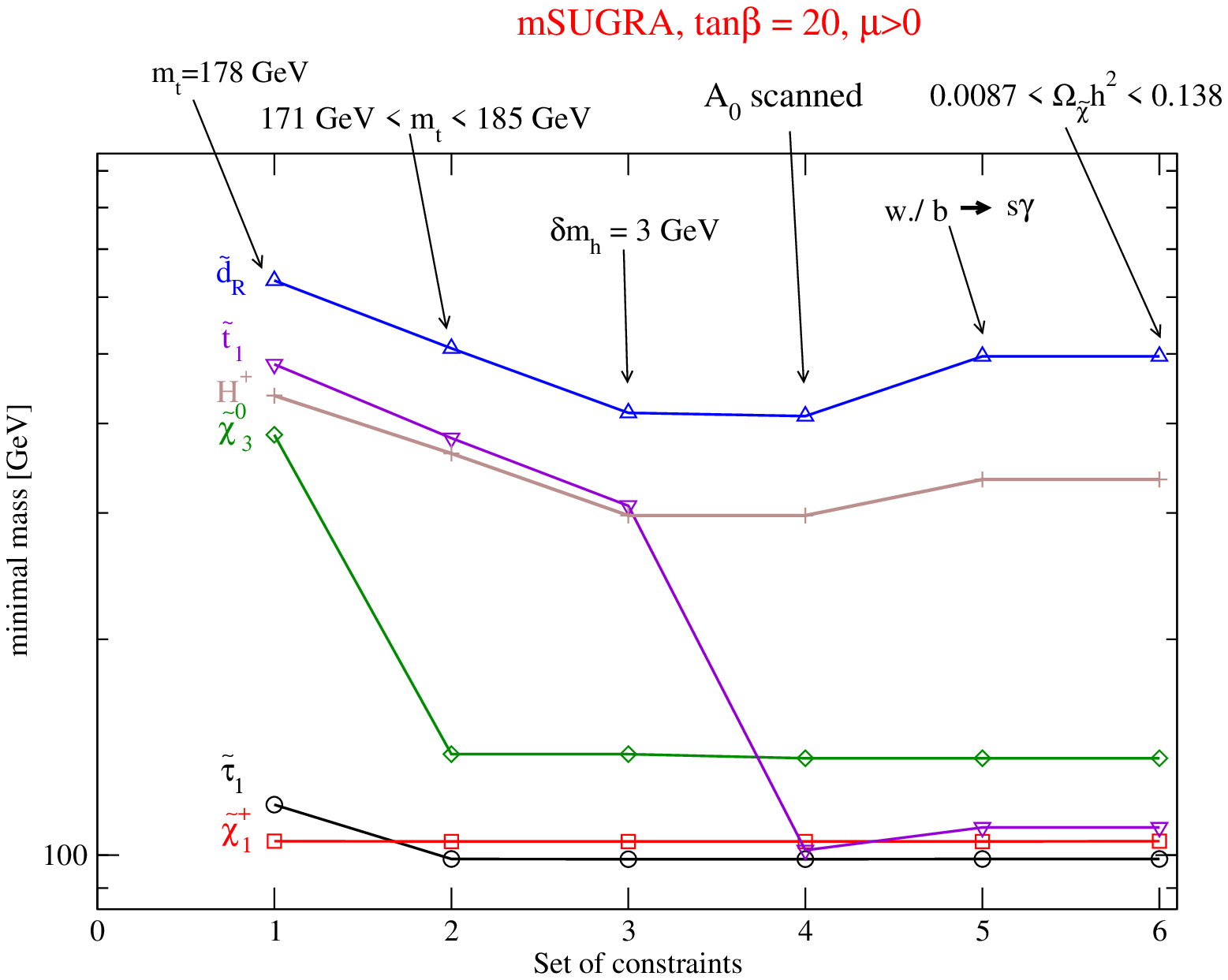}
\end{minipage}
\caption{(a) The $\pi^{\pm}$ energy spectrum from signal (solid line) and
  background (dotted line); from  \cite{Boos:2005ca}. 
(b) The normalized distributions in the fraction of
  $\tau$-jet momentum carried by the  
charged track for $P_\tau$=
+1, +1/2, $-$1/2 and $-$1, and $p^T_{\tau-jet} >$25 GeV; from
\cite{Godbole:2004mq}.  
  (c) Stop discovery reach in $\tilde{t}_1\to c\tilde{\chi}^0_1$ channel in
  the $m_{\tilde{t}_1}$-$m_{\tilde{\chi}^0_1}$ plane -- the 
  $t\to\tilde{t}_1\tilde{\chi}^0_1$ channel was not studied; from
  \cite{milstene}.  (d) Effects of experimental constraints on minimal allowed
  masses of superparticles; from \cite{kneur}. 
 \label{fig:exp}}
 \end{figure}

 Polarisation is a very powerful tool not only for preparing the desirable
 initial state, but also as a diagnosis tool of final states. At this meeting
 improvements of exploiting information on polarisation of fermions coming
 from decays of scalar particles (Higgs, sfermions) have been reported
 \cite{Boos:2005ca,Godbole:2004mq}.  
The charged Higgs boson couples strongly to the fermions
 of the third generation. If the charged Higgs is rather light
 ($M_{H^{\pm}}<M_t$), it decays dominantly to a tau lepton and neutrino
 ($H^{\pm} \to \tau^{\pm} \nu$) making its mass reconstruction impossible.
 The main background comes from decays $W^\pm \to \tau^\pm \nu$.
 However, different structure of $H^{\pm}$ and $W^{\pm}$ electroweak
 interactions implies different tau polarization\footnote{The importance of
   tau polarisation in analyses of stau decays has been
   stressed in Ref.~\cite{polarisation}.  }, which is reflected in
 the energy spectra of the $\tau^{\pm}$ decay products, see e.g.
 fig.~\ref{fig:exp}(a).  
In \cite{Boos:2005ca} detailed computations and Monte Carlo
 simulations for three different sets of MSSM parameters showed that a fit to
 the energy spectrum of the pion in the $\tau^{\pm}\to\pi^{\pm}\nu$ channel
 allows one to infer $M_{H^{\pm}}$ with an uncertainty at the level of
 $0.5$--$1$~GeV.  In \cite{Godbole:2004mq}  the prospects of using $\tau$
 polarisation to probe the composition of $\tilde{\chi}^0_1$ from 
$\tilde\tau^+_1 \tilde\tau^-_1$ production at the ILC 
followed by $\tilde\tau^\pm_1 \rightarrow
 \tau^\pm \tilde{\chi}^0_1$ decay 
 have been investigated. The $\tau$ polarization measurement, via its
 inclusive 1-prong hadronic decay,  can discriminate  SUSY models with
 different  gaugino/higgsino decomposition of the
 neutralino. Fig.~\ref{fig:exp}(b) shows the distribution of the 
 fraction of the visible $\tau$-jet
momentum carried by the charged prong,
$R = p_{\pi^\pm}/p_{\tau-{\rm jet}}$, for four different theoretical models
 (for details we refer to \cite{Godbole:2004mq}).  

Progress on experimental analyses of stop quarks with 
small stop-neutralino mass difference has also been reported to this meeting
\cite{milstene,nowak}. Such analyses are motivated by the stop-neutralino
co-annihilation scenario consistent with relic density and EW baryogenesis,
see next section. 
With small mass difference, the stop decays into neutralino and charm making
the analysis very demanding. Nevertheless, it was shown 
that with the linear collider the region of
co-annihilation down to mass differences $\sim {\cal{O}}$(5 GeV) can be
covered, fig.~\ref{fig:exp}(c), and 
the parameters can be determined accurately enough to reach
precisions for the dark matter predictions comparable  to that from direct
WMAP measurements \cite{milstene}. Additional improvement  in the analysis 
of this scenario can be
obtained if charm triggering were available \cite{nowak}. 

\subsection{Cosmology connection}

Understanding the nature of dark matter is one of the most important
challenges of both particle and astroparticle physics, and collider
experiments may prove useful tool in solving the dark matter puzzle.  Among
many scenarios, weakly interacting massive particles (WIMPs), with masses and
interaction cross sections characterized by the weak scale, are very
compelling and WIMPs appear naturally in low energy supersymmetry models with
$R$-parity. In contrast, the origin of the matter-antimatter asymmetry is more
uncertain.  Electroweak baryogenesis provides a possible scenario 
that relies only on
weak scale physics in which the lightest stop mass must be smaller than the
top quark mass and heavier than about 120~GeV \cite{Balazs:2004bu}. 
Moreover, in this case  the Higgs boson involved in the
electroweak symmetry breaking mechanism must be lighter than 120~GeV, 
and if the lightest neutralino is the dark matter
particle, it must be lighter than stop. This case includes the stop-neutralino
co-annihilation region with acceptable relic density, in which the
stop-neutralino mass difference is small presenting a challenge for stop
searches at colliders \cite{milstene,nowak}.  

Other cosmologically variable regions of parameter space have been discussed at
length in the Cosmological Connections Working Group \cite{CCWG}. 
In our
session an update of all experimental
constraints, including dark matter, on the mSUGRA models and prospects for
superparticle production at the ILC has been presented 
\cite{kneur}.  Special emphases has been put on determining minimal allowed
superparticle masses, which might be more interesting than the ``size'' of
allowed parameter space.  It is found that mSUGRA models with thermal
$\tilde{\chi}^0_1$ as a dark matter particle work fine and that lower bounds
on superparticle masses are only very mildly affected by the DM constraints,
see fig.~\ref{fig:exp}(d). As a result, a possibility of copious superparticle
production even at the first stage of ILC remains open.

 \begin{figure}
(a) \hspace{5cm} (b) \hspace{3cm} \phantom{a}\\
\begin{minipage}[b]{5cm}
\includegraphics[height=3cm,width=3cm]{kilian.ps}~~~
\par\vspace{2mm}
\end{minipage}
\begin{minipage}[t]{5cm} 
\includegraphics[height=3.5cm,width=3.5cm,angle=90]{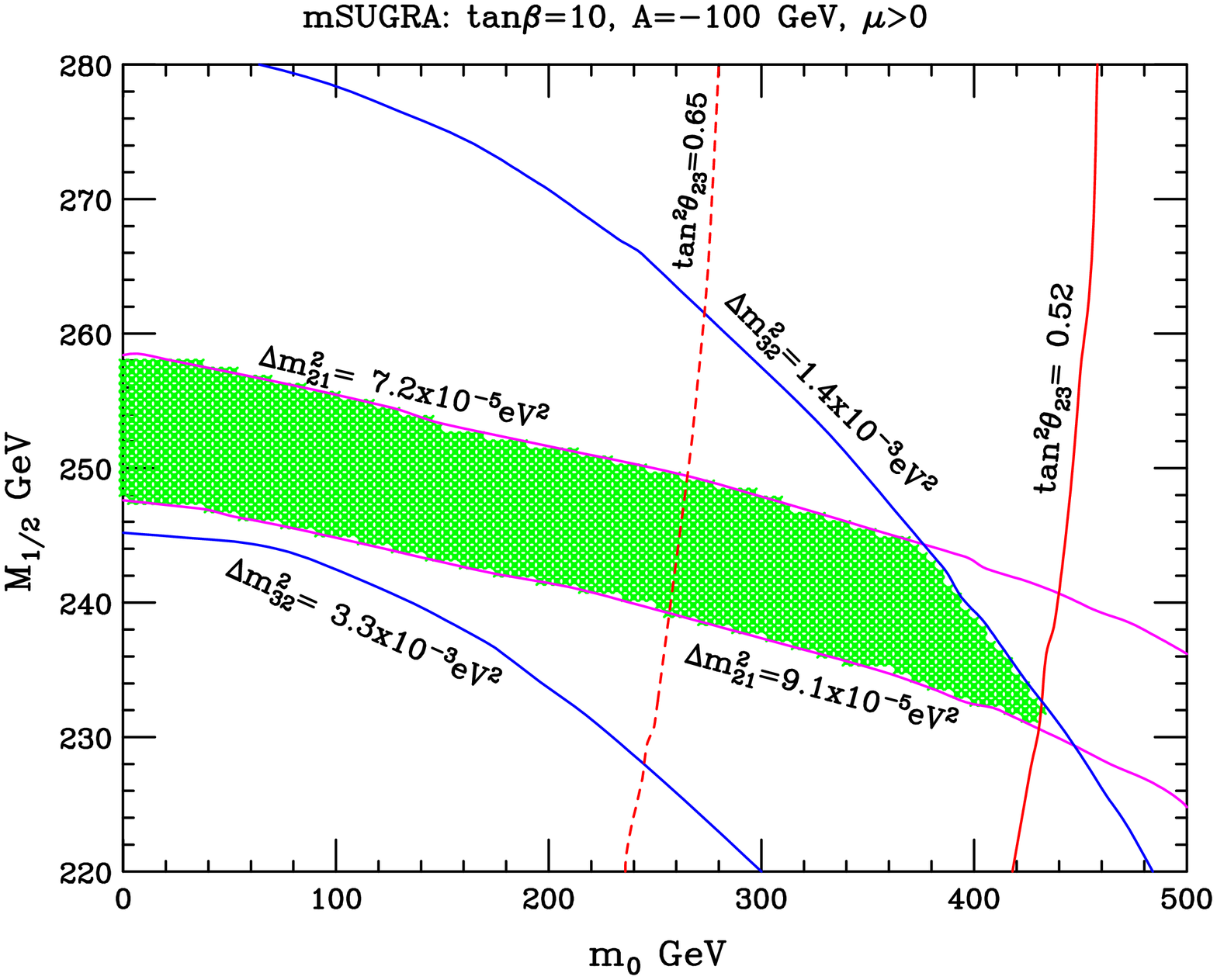}
\end{minipage}
\caption{(a) RGE flow of the anomalous chargino mixing parameters; from 
\cite{Kilian:2005kr}. (b)
  $m_0$-$m_{1/2}$ region with 
  solutions to neutrino physics passing all experimental constraints; from  
\cite{Diaz:2005ii}. 
 \label{fig:beyond}}
 \end{figure}

\subsection{Beyond the MSSM}

The MSSM is  defined as  
an effective low energy model with a)~minimal particle content, 
b)~$R$-parity conservation, 
c)~most general soft supersymmetry breaking terms, which however provide a
technical solution to the hierarchy problem. 
At the meeting several presentations went beyond this minimal set of
assumptions. 

Recently an idea of splitting the
supersymmetry-breaking scale between the scalar and the gaugino
sector has been proposed \cite{sS}. By arranging 
squarks and sleptons very heavy (somewhere between
several tens or hundreds TeV and the GUT scale) with charginos and
neutralinos 
at the TeV scale or below, 
dangerous flavor-changing neutral current transitions, electric dipole
moments, and spurious proton-decay operators can be eliminated, however at a
price of fine-tuning the Higgs sector. In this case the 
low-energy effective theory is quite simple with a SM-like Higgs boson, 
four neutralinos, two charginos, and a gluino. Since squarks are very heavy,
the gluino is long-lived providing a clear signature. Also   the neutralino and
chargino Yukawa couplings deviate from their usual MSSM prediction since they
evolve differently below the splitting scale $\tilde m$. 
Fig.~\ref{fig:beyond}(a) shows the evolution of four anomalous 
chargino mixing parameters (for $\tilde m=10^9$ GeV). By combining
precision measurements of chargino and neutralino masses and Yukawa couplings
at the ILC with gluino mass and life-time at the LHC, it has
been demonstrated \cite{Kilian:2005kr} that the nature of the model can be
verified. 

The simplest extension of the MSSM includes
an additional singlet superfield $S$. It is motivated by the $\mu$ problem of
MSSM since the vacuum expectation value of the scalar
component of $S$ provides an effective $\mu$. The spectrum, apart from
the sparticle content of MSSM, includes two additional Higgs scalars and a 
neutralino which can mix with
the four neutralinos of the MSSM. The parameters of the nMSSM can conspire to
mimic the mass spectrum of the first four neutralinos. The question then arises
how to differentiate experimentally MSSM from nMSSM if at the first stage of
ILC with $\sqrt{s}=500$ GeV 
only two light neutralinos and light chargino are seen. It has been shown
in a particular scenario 
\cite{Moortgat-Pick:2005vs}  that precision measurements of light chargino
and 
neutralino masses at ILC with LHC data on heavier neutralino
$\tilde{\chi}^0_3$  can point to an inconsistent set of SUSY parameters 
under a
working hypothesis of MSSM.    By a moderate rise of the collider energy to
650 GeV the MSSM hypothesis can be falsified and the true nature of nMSSM
revealed.  

 \begin{figure}
(a) \hspace{4cm} (b) \hspace{4cm} (c)\\
\begin{minipage}[b]{5cm}
\includegraphics[width=4cm,height=3.8cm,angle=-90]{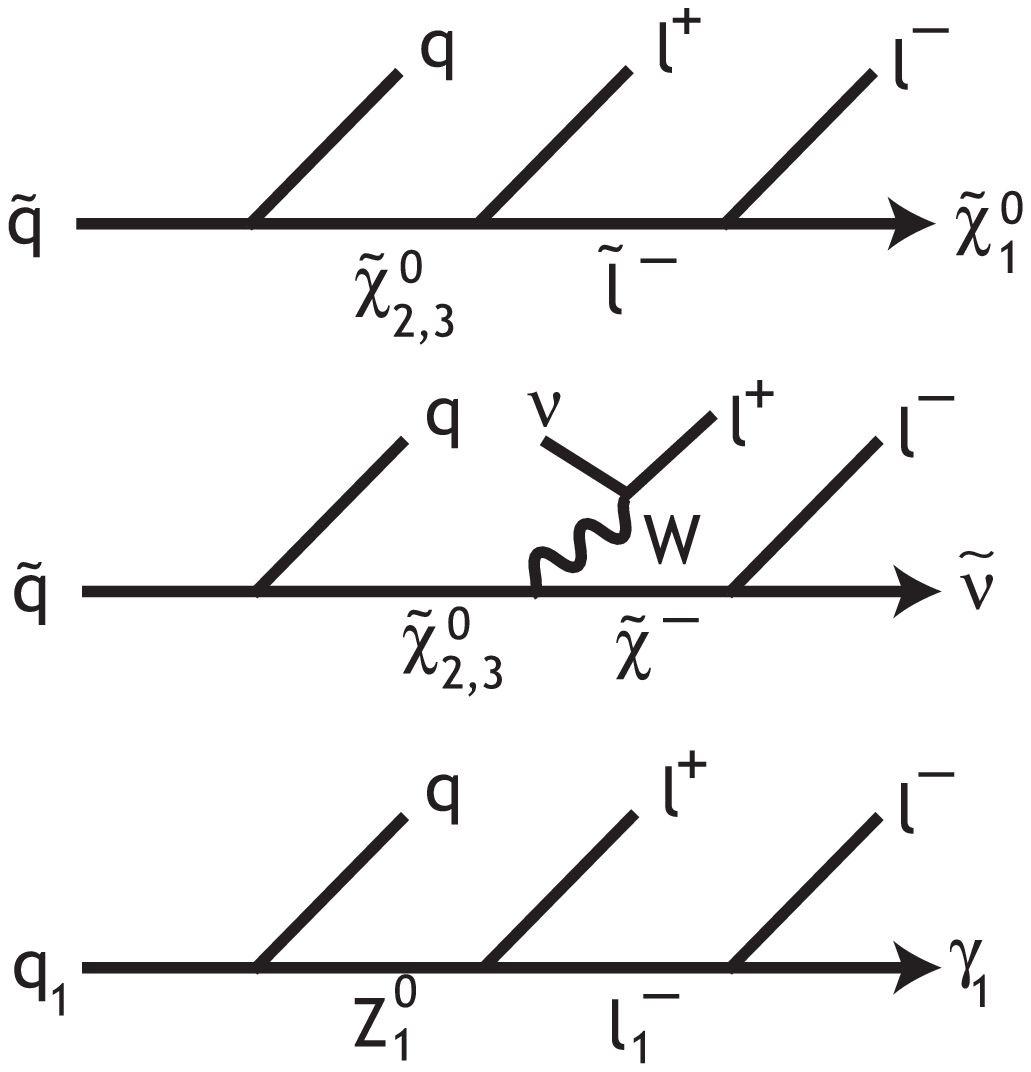}%
\par\vspace{2mm}
\end{minipage}
\begin{minipage}[t]{10cm} 
\includegraphics[width=4cm,height=4cm]{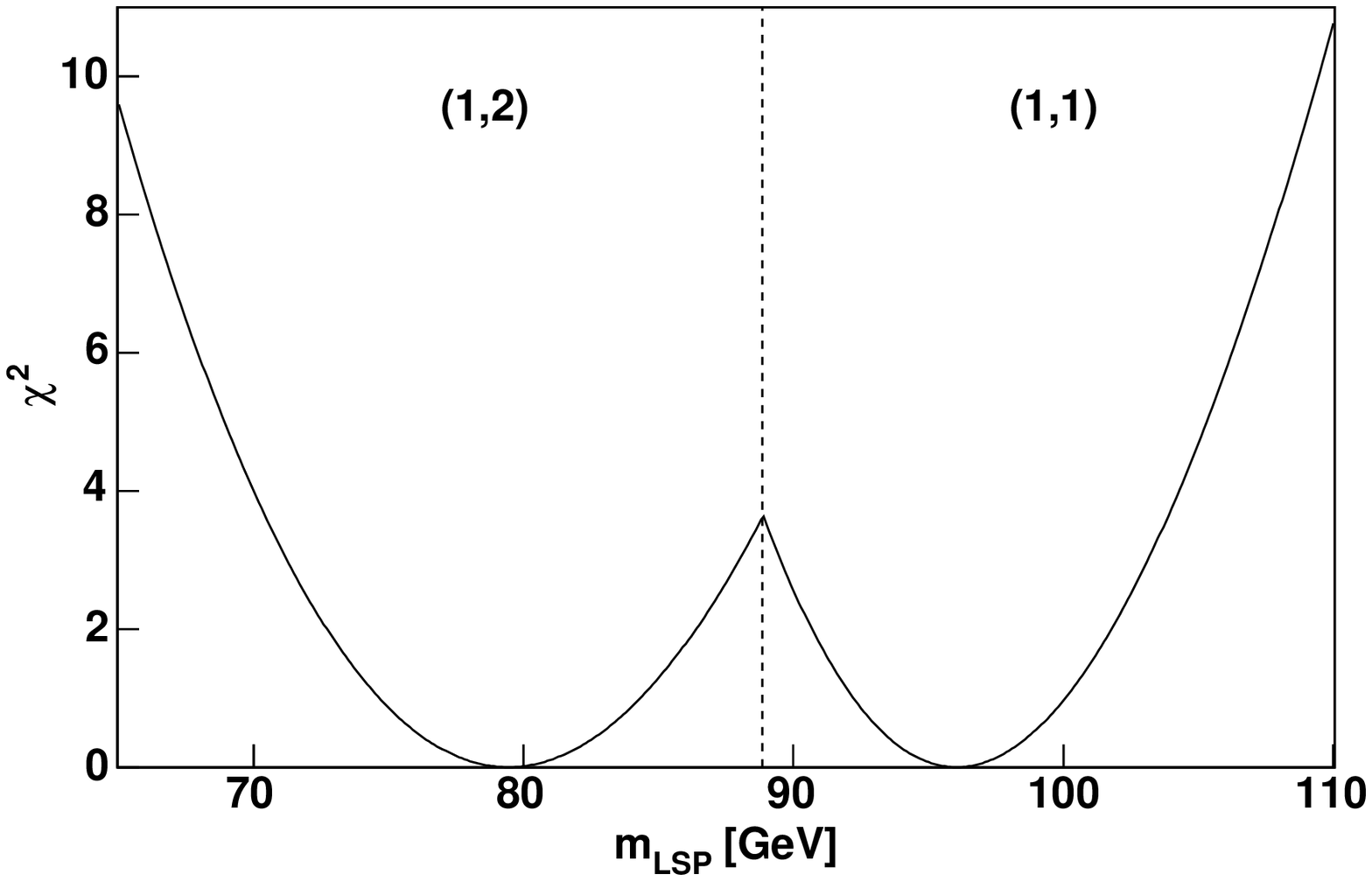}~~~~~~%
\includegraphics[width=4.2cm,height=4.3cm]{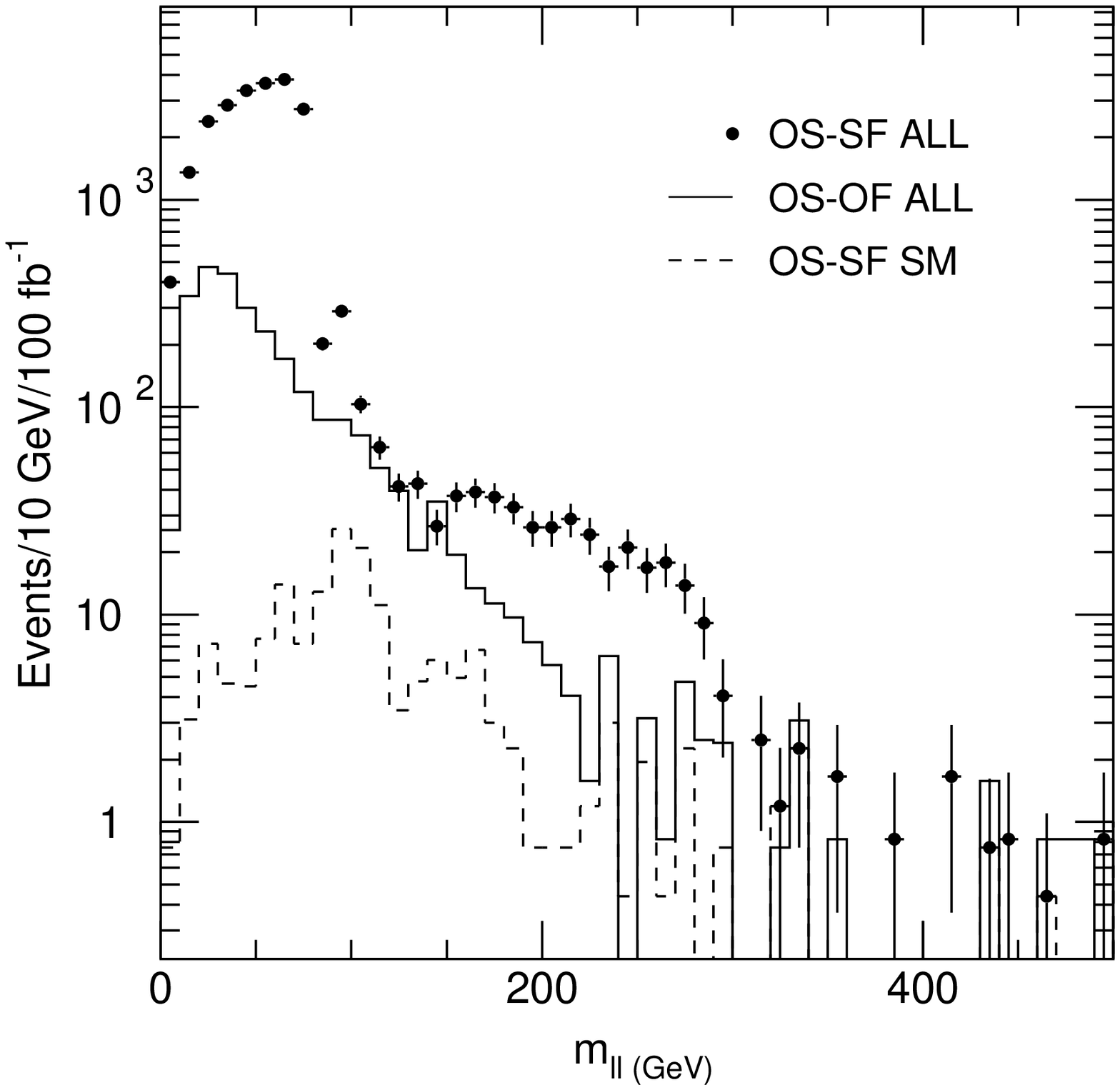}%
\end{minipage}
 \caption{(a) (S)quark decay chains; figure from \cite{peskin}. 
(b) $\chi^2$ as a function of
 $m_{\tilde{\chi}^0_1}$; from \cite{Gjelsten:2005sv}. 
(c) Invariant dilepton mass spectrum
for the decay chain. 
The OS-SF dilepton edge from heavy neutralino is between 
200~GeV and 400~GeV; from \cite{Desch:2003vw}. \label{fig:chain}}
 \end{figure}

\newcommand{\rpv}{\slash\hspace{-2.5mm}{R}_{p}}

In the MSSM 
R--parity implies that 
particles have $R_p$=+1 and their superpartners $R_p$=$-$1. Since  $R_p$ is
conserved, the  LSP must be stable and becomes a candidate for dark
matter particle. However,  R--parity
conservation has no strong theoretical justification since the
superpotential admits explicit R--parity violation ($\rpv$). At this meeting
new developments in models with  explicit bilinear R--parity breaking 
have been presented \cite{Diaz:2005ii}. 
The bilinear terms $  \epsilon_i \hat{L}_i\hat{H}_u$ in the superpotential    
($\hat{H}_u,\hat{L}$ are the Higgs and left--handed lepton superfields) induce
mixing between the neutralinos and neutrinos, forming a $7\times7$ mass 
matrix \cite{Diaz:1997xc}.
Thus the model  provides a 
framework for generating neutrino masses and mixing angles, but at the expense
that the LSP is no
longer stable. For example, fig.~\ref{fig:beyond}(b) shows the $m_0$-$m_{1/2}$
parameter space in the $\rpv$-mSUGRA model consistent with neutrino physics.  
The model leads to definite predictions for the branching ratios of
superparticles which can be tested at future colliders.

Gauge theories with extra dimensions provide an exciting possibility of
unifying gauge and Higgs fields. 
The higher dimensional components of gauge fields
become scalar fields below the compactification scale and  
are identified with 
the Higgs fields in the gauge-Higgs unification theory.
Through quantum corrections, 
the Higgs can take a vacuum expectation value,
and its mass is induced, however 
the radiatively induced mass tends to be small. 
Exploiting  useful expansion formulae for
the effective potential it was shown   \cite{Haba:2004bh} that 
even a small number of bulk field
can generate the suitable heavy Higgs mass. 
The case of introducing 
the soft SUSY breaking 
scalar masses in addition to
the Scherk-Schwarz SUSY breaking and obtaining the heavy Higgs
mass due to the effect of the scalar mass has also been discussed.


\subsection{LHC/ILC Connection}
If a low-energy SUSY scenario with
squarks/gluinos below $\sim$ 2-3 TeV is realized, the LHC will see SUSY
signals. Many channels from squark and/or gluino decays may be observed and
measured. If the decay chain is long enough, like the first squark decay chain
shown in fig.\ref{fig:chain}(a), 
from a knowledge of kinematical endpoints, in
particular those of the  invariant mass distributions $m_{qll}$,
$m_{ql(low)}$,  $m_{ql(high)}$   and $m_{ll}$, 
the masses of the unstable particles can be reconstructed
\cite{Baer:1995va}.  Indeed, the 
endpoints of these distributions can be 
expressed explicitly in terms of $m_{\tilde{q}}$, $m_{\tilde{\chi}^0_{2,3}}$,
  $m_{\tilde{\chi}^0_{1}}$ and $m_{\tilde{l}}$. 
However, 
ambiguities in the masses extracted from endpoint measurements can occur 
(even when experimental
uncertainties are neglected) since the kinematical endpoints are composite
functions of the unknown masses.  This is shown in fig.\ref{fig:chain}(b)
where the false minimum around 80 GeV is found \cite{Gjelsten:2005sv}.
Moreover, 
the same visible final state $l^+l^-q$ can be generated by a different
SUSY decay chain  (second decay chain in fig.\ref{fig:chain}(a)), or
a Kaluza-Klein  
decay chain in a non-supersymmetric model with universal extra dimensions
\cite{Cheng:2002ab} 
(last decay chain in fig.~\ref{fig:chain}(c)). Therefore, the correlation  of
measured masses with particles may not be unique. What differs the decay
chains is the spin of intermediate particles. Although the LHC might  have
some   
sensitivity to spin \cite{Barr:2004ze}, 
both problems -- spin and mass ambiguities -- can be
disentangled with the help of the ILC 
by verifying masses and 
measuring spins of new particles \cite{Weiglein:2004hn}.
In addition, the precise  measurements of kinematically accessible sparticles
at the ILC can facilitate the correct interpretation of the kinematical
endpoints, like the  largest 
observed edge in the opposite-sign same-flavor (OP-SF) dilepton invariant mass
spectrum $m_{ll}$ shown in fig.\ref{fig:chain}(c) as originating from
heavy neutralino in the MSSM context \cite{Desch:2003vw}.

 \begin{figure}
(a) \hspace{4cm} (b) \hspace{4cm} (c)\\
\begin{minipage}[b]{5cm}
\includegraphics[width=4cm,height=5cm,angle=-90]{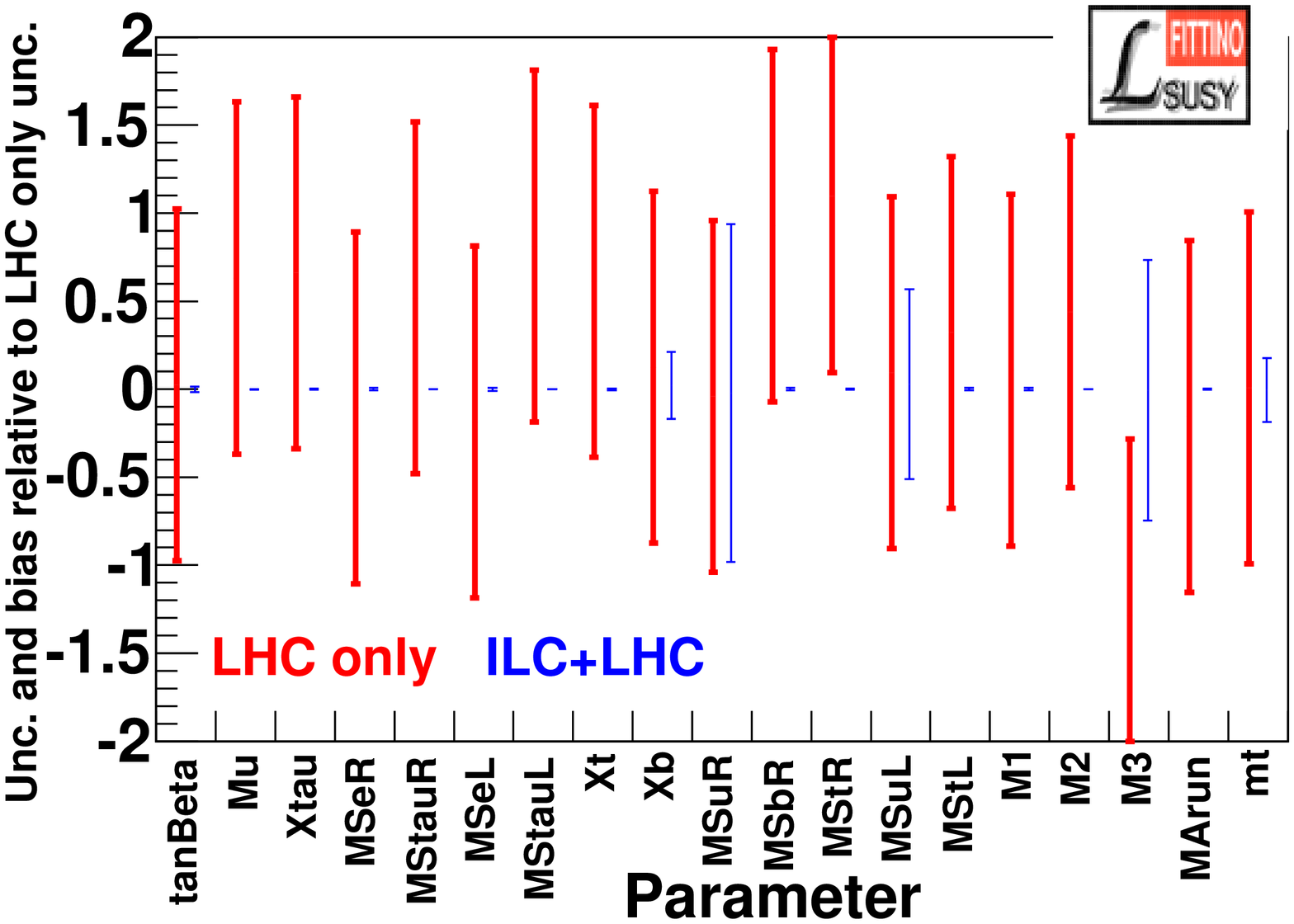}~~~~%
\par\vspace{2mm}
\end{minipage}
\begin{minipage}[t]{11cm} 
\includegraphics[width=5cm,height=4cm]{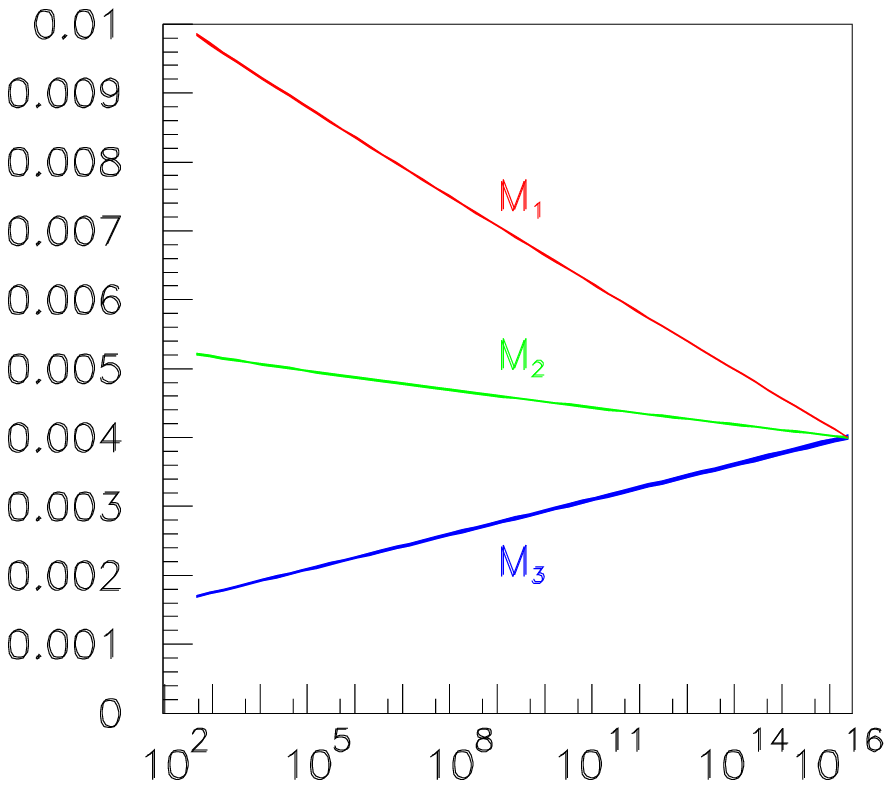}~%
\includegraphics[width=5cm,height=4cm]{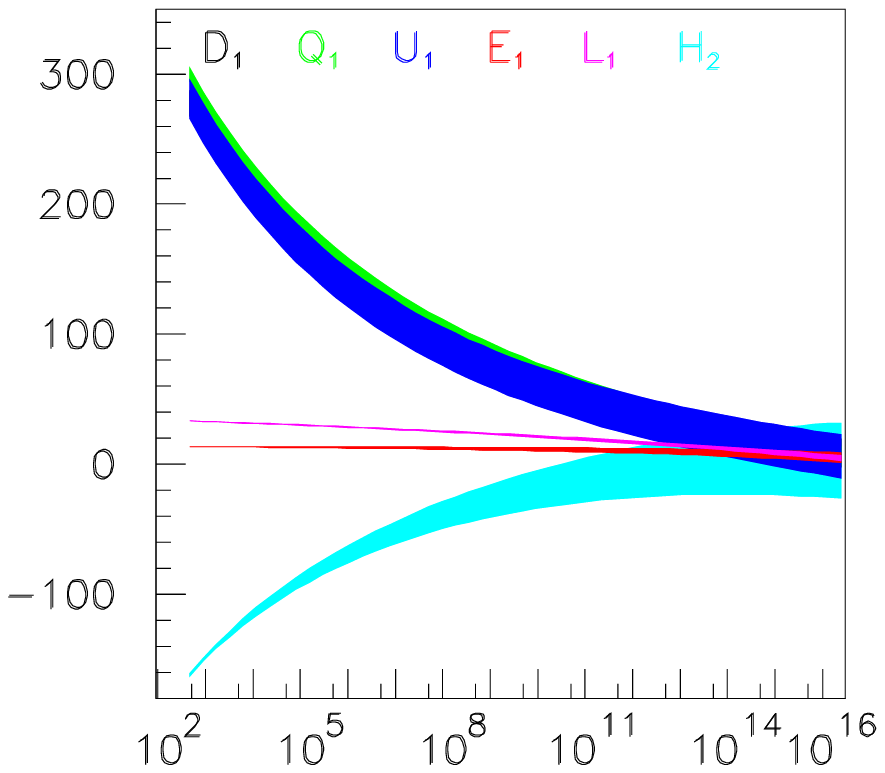}%
\end{minipage}
 \caption{(a) Relative errors on SUSY parameters 
from LHC data and LHC+ILC; from \cite{Bechtle:2005ns}.  
 Evolution of  gaugino  (b) and scalar (c) mass parameters; 
from \cite{Blair:2002pg}.
\label{fig:fit}}
 \end{figure}

Another cosmology-motivated topic is the sensitivity to heavy sfermion 
masses \cite{deboer} that are 
beyond the collider kinematic reach.  At the ILC the lepton forward-backward
asymmetry  in chargino-pair production can provide interesting constraints on
virtual sleptons mediating the chargino decay \cite{Moortgat-Pick:2000uz}.
Preliminary studies of mSUGRA points 
presented at this meeting show that at the LHC the use of a
sophisticated   Kolmogorov-Smirnov test for the dilepton mass spectra from the
neutralino decay can at least distinguish the high-mass
from low-mass scenarios and  a lower limit on the universal scalar 
mass parameter $m_0$ can be determined \cite{Birkedal:2005cm}

Many 
interesting channels can be exploited to extract the basic
supersymmetry parameters when combining experimental information from
sharp edges in mass distributions at LHC with measurements of decay
spectra and threshold excitation curves at an $e^+e^-$ collider with
energy up to 1 TeV. 
From the simulated experimental  data with their errors
available  global analysis programs 
\cite{Lafaye:2004cn} can  exploit  coherently masses, cross sections,
branching ratios etc,  
to extract the
Lagrangian parameters.  The present quality
of such an analysis for the SPS1a' scenario can be judged from  
fig.~\ref{fig:fit}(a) where the relative errors on parameter
determination from LHC data and LHC+ILC are shown \cite{Bechtle:2005ns}.
With the parameters extracted at the scale $\tilde{M}$, 
the reconstruction of the fundamental supersymmetric theory
and the related microscopic picture of the mechanism breaking 
supersymmetry can be investigated  
in the bottom-up approach \cite{Blair:2002pg}
in which the extrapolation from
$\tilde{M}$ to the GUT/Planck scale is performed by the renormalization group
evolution for all parameters.  Typical examples
for the evolution of the gaugino and scalar mass parameters are
presented in fig.~\ref{fig:fit}(b,c) \cite{SPA}.
A comprehensive account of SUSY studies within the LHC/ILC context can be
found in Ref.~\cite{Weiglein:2004hn}.

\section{SUMMARY}

Much progress has been achieved during the last year since LSCW04.
On the theory side many higher-order calculations have been completed and
implemented in numerical codes. The SPA Convention and Project has been
launched which should prove very useful in streamlining discussions and
comparisons of different calculations and experimental analyses. 
On the experimental side many  analyses are still 
based on lowest--order expressions.  In our future studies it is 
important to ensure that new
information from ILC should both significantly 
improve accuracies of SUSY studies at the LHC, as
well as permit calculation of dark matter density to check
cosmology/astrophysics measurements. 
Case  studies have shown that 
a high luminosity ILC with
polarised beams, and with additional $e\gamma$, $\gamma\gamma$ and
$e^-e^-$ modes, can provide high quality data for the precise 
determination of
low-energy SUSY Lagrangian parameters and  
with  LHC data  in the bottom--up approach, 
through the evolution of the parameters from the electroweak scale,  
can reveal the regularities  at  high scales.

\begin{acknowledgments}
Work supported by the Polish KBN Grant  2 P03B 040 24 (for years 2003-2005) 
and 115/E-343/SPB/DESY/P-03/DWM517/2003-2005. I would like to thank S.Y.~Choi,
H.U.~Martyn, M.~Peskin, P.M.~Zerwas for many suggestions, and 
all contributors to the SUSY 
Working Group for their efforts to make the workshop
a success. Hospitality of the Theory Group at SLAC, where this write-up has
been finalised, is greatly acknowledged.  

\end{acknowledgments}

\end{document}